\documentclass[twocolumn]{aastex63}
\usepackage[normalem]{ulem}  
\usepackage[T1]{fontenc}
\usepackage{ae,aecompl}
\usepackage{graphicx}
\usepackage{amsmath}
\usepackage{amssymb}
\usepackage{supertabular}
\usepackage{hyperref}
\usepackage{xcolor}
\usepackage{academicons}

\def\plb{Physics Letters B}
\def\bea{\begin{eqnarray}} \def\eea{\end{eqnarray}}
\def\fm3{\;\text{fm}^{-3}}

\newcommand{\Msun}{\,M_{\odot}}
\newcommand{\km}{\hbox{$\,{\rm km}$}}

\begin{document}
\title{Revisiting the post-glitch relaxation of the 2000 Vela glitch with the neutron star equation of states in the Brueckner and relativistic Brueckner theories}

\correspondingauthor{Ang Li}
\email{liang@xmu.edu.cn}

\author{Xinle Shang}
\affiliation{Institute of Modern Physics, Chinese Academy of Sciences, Lanzhou 730000, China}

\author{Ang Li}
\affiliation{Department of Astronomy, Xiamen University, Xiamen, Fujian 361005, China}

\date{\today}

\begin{abstract}
We revisit the short-term post-glitch relaxation of the Vela 2000 glitch in the simple two-component model of pulsar glitch by making use of the latest realistic equations of states from the microscopic Brueckner and the relativistic Brueckner theories for neutron stars, which can reconcile with the available astrophysical constraints.
We show that to fit both the glitch size and the post-glitch jumps in frequency derivatives approximately one minute after the glitch, the mass of the Vela pulsar is necessarily small, and there may be demands for a stiff equation of state (which results in a typical stellar radius larger than $\sim\negthickspace12.5\,\rm km$) and a strong suppression of the pairing gap in the nuclear medium.
We discuss the implications of this result on the understanding of pulsar glitches.
\end{abstract}


\section{Introduction}\label{sec:intro}

Glitches, i.e., the sudden spin-up of pulsars with subsequent recovery, were discovered during pulsar timing studies in the Vela pulsar (PSR J0835-4510 or PSR B0833-45) more than fifty years ago~\citep{1969Natur.222..228R}.
They are generally agreed to be the result of the complicated interplay between the superfluid component and the rest of the star~\citep{1969Natur.224..673B,1975Natur.256...25A}, i.e., the two-component framework.
Glitch observations are important for understanding the properties of neutron stars and provide a unique way to probe neutron star internal structure and dynamics.
Many unknown questions concerning both the trigger and the relaxation of a glitch are closely related to the microphysical state of the matter and hence to the equation of state (EOS) and the structure of the neutron stars.

With a total of 22 glitches in about 50 years, the Vela pulsar is one of the most frequently glitching pulsars of over 2000 known pulsars. 
Among the many efforts made to try to unravel the glitches in Vela-like pulsars, it has been shown that the sizes and subsequent recoveries can be understood to some extent from the hydrodynamic evolution of the vortices~\citep{2009MNRAS.400.1859S,2010MNRAS.405.1061S,2010MNRAS.409.1253V,2011ApJ...743L..20P,2012MNRAS.420..658H,2012MNRAS.427.1089S,2016MNRAS.460.1201H,2018MNRAS.481L.146H,2020MNRAS.496.2506G}; furthermore, a ``snowplow'' model has been proposed~\citep{2011ApJ...743L..20P}, which incorporates microscopically based pinning forces throughout the stellar crust.
\citet{2012MNRAS.427.1089S} extended the snowplow model from a polytropic EOS to realistic EOSs obtained in the popular relativistic mean-field (RMF) model.
They found that the model could reproduce the main features of giant glitches in the Vela pulsar, including the size and post-glitch jump infrequency derivatives of the January 2000 glitch~\citep{2002ApJ...564L..85D}. Note that a fourth exponential decay term (approximately 1 minute) was found, which represents the fastest decay term to date.
\citet{2013ApJ...764L..25H} further extended the model to account for possible pinned vorticity in the core due to the interaction between vortices and flux tubes.

In those previous studies on short-term transient behaviours around glitch epoch, the proton fraction was adopted from a microscopic EOS in the Brueckner theory~\citep{2004PhLB..595...44Z}, which is different from the RMF EOSs used for the pulsar.
Due to the critical dependence of the star composition (including the proton fraction) on the EOS, in the present study, we intend to revisit the 2000 Vela glitch based on a consistent determination of both the EOS and the proton fraction from the microscopic (relativistic) Brueckner theory.
Formerly, consistent EOS and composition on the basis of the Skyrme nuclear effective force have been adopted by \citet{2018MNRAS.475.5403A} for the study of the general relativistic corrections to pulsar glitch amplitudes.

The Brueckner theory has been proven to be one of the most advanced microscopic theories~\citep{book}, as it allows an $ab~initio$ calculation of neutron star EOS from relativistic nucleon-nucleon (NN) potentials.
In particular, the exact solution of the Brueckner-Bethe-Goldstone equation has been made available recently, which is free from the total momentum approximation that was commonly adopted in previous works~\citep[see more discussions in][]{2021PhRvC.103c4316S}.
Furthermore, high-precision NN potentials (i.e., the pvCD-Bonn potentials) that are applicable in the relativistic Brueckner theory~\citep{rbhf} have been generated recently and can precisely describe the charge dependence in NN scattering data~\citep[see more discussions in][]{2019ChPhC..43k4107W}.
The present calculations are done from both the Brueckner and relativistic Brueckner theories.

The aim of this paper is to better understand the microscopy and macroscopy of a neutron star with glitch observations.
In particular, we aim to examine to what extent a simple two-component model of glitch 
can be applicable to observational constraints from the Vela pulsar.
Section 2 is dedicated to formulating the two-component glitch model.
Section 3 demonstrates the inner structure of a pulsar based on the microscopic EOSs within the Brueckner and relativistic Brueckner theories.
Section 4 presents an application of the EOS and the glitch model to the Vela pulsar.
Finally, we present a discussion and the conclusions in Section 5.

\section{The two-component model}\label{sec:2model}

In the simple two-component framework, one aspect is the charged component (labelled as index ``p''), and the other is the superfluid component (labelled as index ``n'').
The visible $\textrm{p}$ component includes crustal lattices, core protons, and neutrons (core neutrons can be partially coupled, as discussed later).
The $\textrm{n}$ component is usually assumed to be crustal neutrons, which may indeed be the case according to recent studies on the type of proton superconducting in the core~\citep{2012MNRAS.420..658H,2013ApJ...764L..25H} and on the interaction between neutron rotational vortices and proton flux tubes~\citep{2009PhRvL.103w1101B}.

A glitch is triggered by the sudden transfer of angular momentum from $\textrm{n}$ to $\textrm{p}$ at some critical spin lag between the two.
When the superfluid vortices are pinned to the crustal lattice, the charged component evolves
according to the following~\citep{2011ApJ...740L..35G}:
\bea
I_\textrm{p} \dot{\Omega_\textrm{p}}  = -a \Omega_\textrm{p}^3 \ ,
\eea
where the term on the right side represents the standard torque due to a magnetic dipole ($a$ is a coefficient).
From balancing the pinning force to the Magnus force, 
one determines the critical spin lag at which the vortices will unpin (as shown in more details later in Section~\ref{sec:step}).
Normally, the lag is small; we then have the following:
\bea
{\Delta \Omega_\textrm{p}  \over \Omega_\textrm{p} } \approx {I_\textrm{n} \over I_\textrm{p}} {\Delta t_{\rm g}\over 2 \tau_\textrm{c}}\ ,
\label{eq2}
\eea
where $\Delta t_{\rm g}$ is the interglitch time ($\sim\negthickspace2$ yr for Vela), and $\tau_\textrm{c} = - { \Omega_\textrm{p} / 2 \dot{\Omega_\textrm{p}}}$ is the characteristic age of the pulsar ($\sim\negthickspace11.3~\rm kyr$ for Vela).

The fractional moment of inertia $I_\textrm{n}/I_\textrm{p}$ is related to the glitch activity $A_\textrm{g}=(1/T)(\sum\Delta\Omega_\textrm{p})/\Omega_\textrm{p}$ as follows:
\bea
2\tau_\textrm{c} A_\textrm{g} 
\lesssim \frac{I_\textrm{n}}{I_\textrm{p}} \ ,
\label{eq:Ag}
\eea
where $T$ is the total data span for glitch monitoring, and $\sum(\Delta\Omega_\textrm{g})$ is the sum of all observed glitch frequency jumps.
The glitch activity $A_\textrm{g}$ can be estimated for systems that have exhibited at least two glitches of similar magnitude, such as the Vela pulsar.
Using the observed value of Vela's activity parameter $A_\textrm{g}=1.91\times10^{-9}$ per day~\citep{2011MNRAS.414.1679E}, 
the glitch observations have put a constraint on the fractional moment of inertia, which is $I_\textrm{n}/I_\textrm{p}\gtrsim  1.6\%$~\citep{1999PhRvL..83.3362L,2012PhRvL.109x1103A,2021Univ....7....8M}.
It has been argued that the ``entrainment'' of the neutron superfluid by the crystalline structure of the crust greatly reduces its mobility; namely, Equation (\ref{eq:Ag}) is modified to the following~\citep{2012PhRvL.109x1103A,2013PhRvL.110a1101C}:
\bea
2\tau_c A_\textrm{g} \frac{\langle m_n^* \rangle}{m_n}\lesssim  \frac{I_\textrm{n}}{I_\textrm{p}} \ ,
\label{eq4}
\eea
where $\langle m_n^*\rangle$ is the average effective neutron mass. $\langle m_n^* \rangle/m_n$ is in the range of $\sim\negthickspace1.5\negthickspace-\negthickspace6$~\citep{2005NuPhA.747..109C,2012PhRvC..85c5801C,2017JLTP..189..328C,2017PhRvL.119f2701W}.

The glitch rise time is usually very short. Presently, the best constraint comes from the 2016 Vela glitch, which is less than $12.6$ seconds~\citep{2019NatAs...3.1143A}. 
Correspondingly, it has been suspected that only a fraction of the core neutrons ($\lesssim30\%$, as seen later in Section~\ref{sec:step}) might be coupled to the charged component of the crust on glitch-rise time scales~\citep{2012MNRAS.420..658H}.
Previously, \citet{2015ChPhL..32g9701L,2016ApJS..223...16L} argued the presence of the glitch crisis in the framework of the microscopic Brueckner theory, namely, if when accounting for the entrainment effect, there is not enough angular momentum transferred to trigger the big Vela-like glitches from only the crustal superfluid neutrons. That is, Equation (\ref{eq4}) cannot be guaranteed.
Such calculations were performed under the assumption that the whole core part is coupled to the charged visible part during a glitch. Here, we relax this assumption and consider a more realistic scenario of a partially coupled core, which equivalently decreases the denominator $I_{\rm p}$ on the right side of Equation (\ref{eq4}). Therefore, the glitch crisis is currently not a problem. That the glitch crisis can  be solved with the core superfluid had been already shown in \citet{2014ApJ..788L..11G,2015IJMPD..2430008H,2015MNRAS.449.3559H}. 

We now present the inner structure of a pulsar based on the modern EOSs within the (relativistic) Brueckner theory and confront the theoretical results in the snowplow model with the only one available observation of the short-term recovery component of the Vela glitch~\citep{2002ApJ...564L..85D}.

\section{Microscopic EOS and stellar structure}\label{sec:bhf}

\subsection{Brueckner-Hartree-Fock (BHF) approach}

The Brueckner theory is based on the re-summation of the perturbation expansion of the ground-state energy of nuclear matter~\citep{book}.
For the present study, the EOS is calculated at the Brueckner-Hartree-Fock (BHF) level and adopts the realistic Argonne $V_{18}$ with a microscopic three-body force~\citep{2002NuPhA.706..418Z}.
The details are described elsewhere~\citep{1999PhRvC..60b4605Z}. Here, we outline it briefly.

In the BHF approach, the infinite series of certain diagrams in perturbation expansion for the total energy of the system is embodied in the so-called $G$-matrix operator, which is adopted instead of the bare nucleon interaction (with care taken for double counting of the same contributions).
The main advantage of the $G$-matrix is that the elements of $G$ do not diverge, as could happen when using a bare nucleon potential. Accordingly, one can start the calculation from a realistic nucleon-nucleon potential with no artificial parameters in the approach.

The $G$-matrix satisfies the Bethe-Goldstone equation:
\begin{eqnarray}
G(\omega)=v +v\frac{Q}{\omega-\epsilon_{1}-\epsilon_{2}+\imath0^{+}}G(\omega)\ ,
\end{eqnarray}
with the particle spectrum in BHF approach:
\begin{eqnarray}
\epsilon(p)=\frac{p^{2}}{2m_{N}}+U(p)\ .
\end{eqnarray}
where $v$ is the bare nucleon potential, and $Q$ is the Pauli operator, which represents the nuclear medium effect. Here we remark the bare nucleon mass as $m_{N}$ and neglect the bare mass difference between neutrons and protons. $U(p)$ is the auxiliary self-consistent potential and is determined by the following:
\begin{eqnarray}
U(p_{1})=
\sum_{p_{2}<p
_{F2}}\textrm{Re}\langle
12|G(\epsilon_{1}+\epsilon_{2})|12\rangle_{A}\ ,
\end{eqnarray}
where $|12\rangle_{A}$ means that the wave function is properly antisymmetrized and $p_{F2}$ denotes the Fermi momentum.
After several self-consistent iterations of Equations $(5)\negthickspace-\negthickspace(7)$, the effective interaction matrix $G$ is obtained. Using this $G$-matrix, the total energy per nucleon can be expressed as follows:
\begin{eqnarray}
E/A=&&\frac{1}{A}\sum_{p<p_{F}}\frac{p^{2}}{2m_N} \nonumber\\
&&+\frac{1}{2A}\sum_{p_{1}<p
_{F1};p_{2}<p
_{F2}}\textrm{Re}\langle
12|G(\epsilon_{1}+\epsilon_{2})|12\rangle_{A}\ .
\end{eqnarray}
Following this, the EOS of uniform nuclear matter, pressure $P$ as a function of the density $\rho$, or energy density $\varepsilon$, can be calculated. The saturation properties and the symmetry energy parameters are found to be consistent with the empirical constraining bands. More results and discussions can be found in \citet{2020PhRvC.101f5801S,2021PhRvC.103c4316S}.

\subsection{Relativistic BHF approach}

Different from the BHF approach, one needs a Dirac spinor instead of the free nucleon wave function (plane wave function) for the description of single particle motion in a nuclear medium within the RBHF approach:
\begin{eqnarray}
u(\emph{\textbf{p}},\emph{\textbf{s}})=\left(\frac{E^{*}_{\emph{\textbf{p}}}+M_{N}^{*}}{2M_{N}^{*}}\right)^{1/2}\left(
\begin{array}{l}
~ ~ ~ ~ ~ ~ 1 \\
\frac{\boldsymbol \sigma \cdot \emph{\textbf{p}}}{E^{*}_{\emph{\textbf{p}}}+M_{N}^{*}}
\end{array}
\right)\chi_{\emph{\textbf{s}}}\ .
\end{eqnarray}
Here, $M_{N}^{*}=m_{N}+U_{S}$ and $E^{*2}_{\emph{\textbf{p}}}=M_{N}^{*2}+\emph{\textbf{p}}^2$. $U_{S}$ represents the scalar potential, and $\chi_{\emph{\textbf{s}}}$ is the Pauli spinor.

In the RBHF calculation, the $G$ matrix, which serves as an effective interaction, can be obtained by solving the in-medium relativistic scattering equation. One of the most widely used covariant scattering equations is the Thompson equation~\citep{1970PhRvD...1..110T}, which is a relativistic three-dimensional reduction of the Bethe-Salpeter equation~\citep{1951PhRv...84.1232S}. In the rest frame of nuclear matter, the Thompson equation takes the following form:
\begin{eqnarray}
G(\emph{\textbf{q}}',\emph{\textbf{q}},\emph{\textbf{P}},\omega)&=&V(\emph{\textbf{q}}',\emph{\textbf{q}})+\int\frac{d\emph{\textbf{k}}}{(2\pi)^3}V(\emph{\textbf{q}}',\emph{\textbf{k}})\frac{M_{N}^{*2}}{E^*_{\emph{\textbf{P}}/2+\emph{\textbf{k}}}E^*_{\emph{\textbf{P}}/2-\emph{\textbf{k}}}}\nonumber\\
&\times& \frac{Q(\emph{\textbf{k}},\emph{\textbf{P}})}{\omega-E_{\emph{\textbf{P}}/2+\emph{\textbf{k}}}-E_{\emph{\textbf{P}}/2-\emph{\textbf{k}}}}G(\emph{\textbf{k}},\emph{\textbf{q}},\emph{\textbf{P}},\omega)\ ,
\end{eqnarray}
where $E_{\emph{\textbf{p}}}$ is the eigenvalue of the Diarc equation in the medium. $\emph{\textbf{P}}=(\emph{\textbf{k}}_{1}+\emph{\textbf{k}}_{2})/2$ and $\emph{\textbf{k}}=(\emph{\textbf{k}}_{1}-\emph{\textbf{k}}_{2})/2$ are the total and relative momentum of the two interacting nucleons with momenta $\emph{\textbf{k}}_{1}$ and $\emph{\textbf{k}}_{2}$, respectively. $\emph{\textbf{q}}$, $\emph{\textbf{k}}$, and $\emph{\textbf{q}}'$ are the initial, intermediate, and final relative momenta of the two nucleons in the medium, respectively. The Pauli operator $Q$ is the same as that used in the BHF approach.

Using this $G$ matrix, the self-energy for the positive energy solution can be calculated as follows:
\begin{eqnarray}
U(1)=\sum_{p_{2}<p
_{F2}}\langle
12|G(E_{1}+E_{2})|12\rangle_{A}\ .
\end{eqnarray}
In addition, the expectation value of the single-particle potential for nucleons with momentum $\emph{\textbf{p}}$ can be expressed by the following:
\begin{eqnarray}
U(\emph{\textbf{p}})=\frac{M_{N}^{*}}{E^{*}_{\emph{\textbf{p}}}}U_{S}+U_{V}\ .
\end{eqnarray}
In general, the scalar and vector potentials $U_{S}$ and $U_{V}$ are momentum dependent; however, this dependence is very weak and can be neglected. The constants $U_{S}$ and $U_{V}$ can be adjusted to the positive energy solution in Equation (11) at the Fermi momentum. Accordingly, the eigenvalues $E_{\emph{\textbf{p}}}$ can be obtained by solving the relativistic Hatree-Fock equation:
\begin{eqnarray}
\{\boldsymbol \alpha \cdot \emph{\textbf{p}} + U_{V} + \beta M_{N}^*\}u(\emph{\textbf{p}})=E_{\emph{\textbf{p}}}u(\emph{\textbf{p}})\ ,
\end{eqnarray}
where $\boldsymbol \alpha=\gamma_{0} \boldsymbol{\gamma}$ and $\beta=\gamma_{0}$ are the Dirac matrices. Equations $(10)\negthickspace-\negthickspace(13)$ should be solved self-consistently by iterating until their convergence. Afterwards, the binding energy per nucleon is evaluated by:
\begin{eqnarray}
E/A&=&\frac{1}{A}\sum_{p_{2}<p_{F2}}\frac{M^*}{E^*}\langle 1 |\boldsymbol \alpha \cdot \emph{\textbf{p}} +  \beta m_{N} |1\rangle \nonumber -m_{N}\\ &+& 
\frac{1}{2A}\sum_{p_{1}<p
_{F1};p_{2}<p
_{F2}}\langle
12|G(E_{1}+E_{2})|12\rangle_{A}\ .
\end{eqnarray}
Then, the EOS of nuclear matter can be calculated in the same manner as that used in the BHF approach. As mentioned in the introduction, in the present RBHF calculations, we adopt charge-dependent Bonn potentials with pseudovector coupling between the pion and nucleon, pvCD-Bonn A~\citep{2019ChPhC..43k4107W}, to evaluate the EOS. 
The saturation properties of nuclear matter for the two employed EOSs in the Brueckner and relativistic Brueckner theories are collected in Table~\ref{t:eos}, together with the constraints from terrestrial experiments~\citep{2006EPJA...30...23S,2010JPhG...37f4038P,2017RvMP...89a5007O,2019EPJA...55..117L,2020PhRvL.125t2702D,2021PhRvL.126q2503R}. 

\begin{table*}
\caption{\label{table:NM} Various saturation properties of two employed nuclear matter EOSs in the Brueckner and relativistic Brueckner theories.
The constraints from terrestrial experiments are $\rho_0\sim\negthickspace0.14\negthickspace-\negthickspace0.17$~fm$^{-3}$, $-B/A\sim\negthickspace15\negthickspace-\negthickspace17$ MeV~\citep{2018PhRvC..97b5805M}, $K = 230 \pm 20$ MeV~\citep{2006EPJA...30...23S,2010JPhG...37f4038P}, $E_{\rm sym} = 31.7 \pm 1.1$ MeV and $L = 59.8 \pm 4.1$ MeV, as reported in \citet{2017RvMP...89a5007O,2019EPJA...55..117L,2020PhRvL.125t2702D}. 
From the new measurement of the updated Lead Radius EXperiment (PREX-II), \citet{2021PhRvL.126q2503R} reported $E_{\rm sym} = 38.2 \pm 4.7$ MeV and $L = 106 \pm 37$ MeV.
Also listed in the two rightmost columns are the maximum gravitational mass of neutron stars ($M_{\rm TOV}$) and the radius of a typical $1.4 \Msun$ star, respectively.}
\setlength{\tabcolsep}{0.8pt}
\renewcommand\arraystretch{1.1}
\begin{ruledtabular}
\vspace{+0.2cm}
\begin{tabular*}{\hsize}{@{}@{\extracolsep{\fill}}ccccccccc@{}}
Model      & $\rho_0$        &   $B/A$    &   $K$  &  $E_{\rm sym}$   & $L$   &  $M_{\rm TOV}$ & $R_{\rm 1.4}$\\
         & (fm$^{-3}$)  &   (MeV)    &   (MeV)  &  (MeV)  &  (MeV)  &  ($\Msun$) & (km )     \\   \hline
BHF     &  0.186       &  $-14.47$  & 213 &  33.0 & 72.8 & 2.20 & 11.09\\
RBHF    &  0.192       &  $-16.83$  & 315 &  36.8 & 80.5  & 2.21 & 12.34\\
\end{tabular*}
\end{ruledtabular} \label{t:eos}
\end{table*}

\subsection{Stellar structure}
To study the structure of neutron stars, we must calculate the composition and EOS of charge-neutral neutron star matter, which consists of neutrons, protons, and leptons ($e^-$, $\mu^-$) in beta equilibrium~\citep{2020JHEAp..28...19L}.
Once the EOS $P(\varepsilon)$ is specified, the stable configurations of neutron stars can be obtained from solving the Tolman-Oppenheimer-Volkoff (TOV) equation~
for pressure $P$ and the enclosed mass $M$:
\begin{eqnarray}
\frac{dP(r)}{dr}&=&-\frac{\mathcal{G}M(r)\varepsilon(r)}{r}\frac{[1+\frac{P(r)}{\varepsilon(r)}][1+\frac{4\pi
r^{3}P(r)}{M(r)}]}{1-\frac{2\mathcal{G}M(r)}{r}},\\
\frac{d M(r)}{dr}&=&4\pi r^{2}\varepsilon(r),
\end{eqnarray}
where $\mathcal{G}$ is the gravitational constant.
The moment of inertia can be calculated under rigid-body rotation in \citet{2011ApJ...743L..20P}:
\begin{eqnarray}
I(r_{1},r_{2})=\frac{8\pi}{3}\int_{r_{1}}^{r_{2}}r^{4}\rho(r)dr.
\end{eqnarray}
Thus, the total moment of inertia is $I_{\rm tot}=I(0,R)$,
with $R$ being the radius of the star.
For later use, $R_{\rm ic}$ is defined as the radius of the inner crust.
The BPS~\citep{1971ApJ...170..299B} and NV~\citep{1973NuPhA.207..298N} EOS has been employed for nonuniform matter present in the outer and inner crust parts of the stars (below $0.08\fm3$ or $\sim\negthickspace0.5\rho_0$), respectively.
That is, the pressures from the BHF/RBHF calculations are replaced by that of \citet{1973NuPhA.207..298N} from the nucleonic density $0.08\fm3$.
We mention here that the NV EOS is based on quantal Hartree-Fock calculations for Wigner-Seitz cells for the inner crust.

\begin{figure*}
\centering
\resizebox*{0.98\textwidth}{0.55\textheight}
{\includegraphics{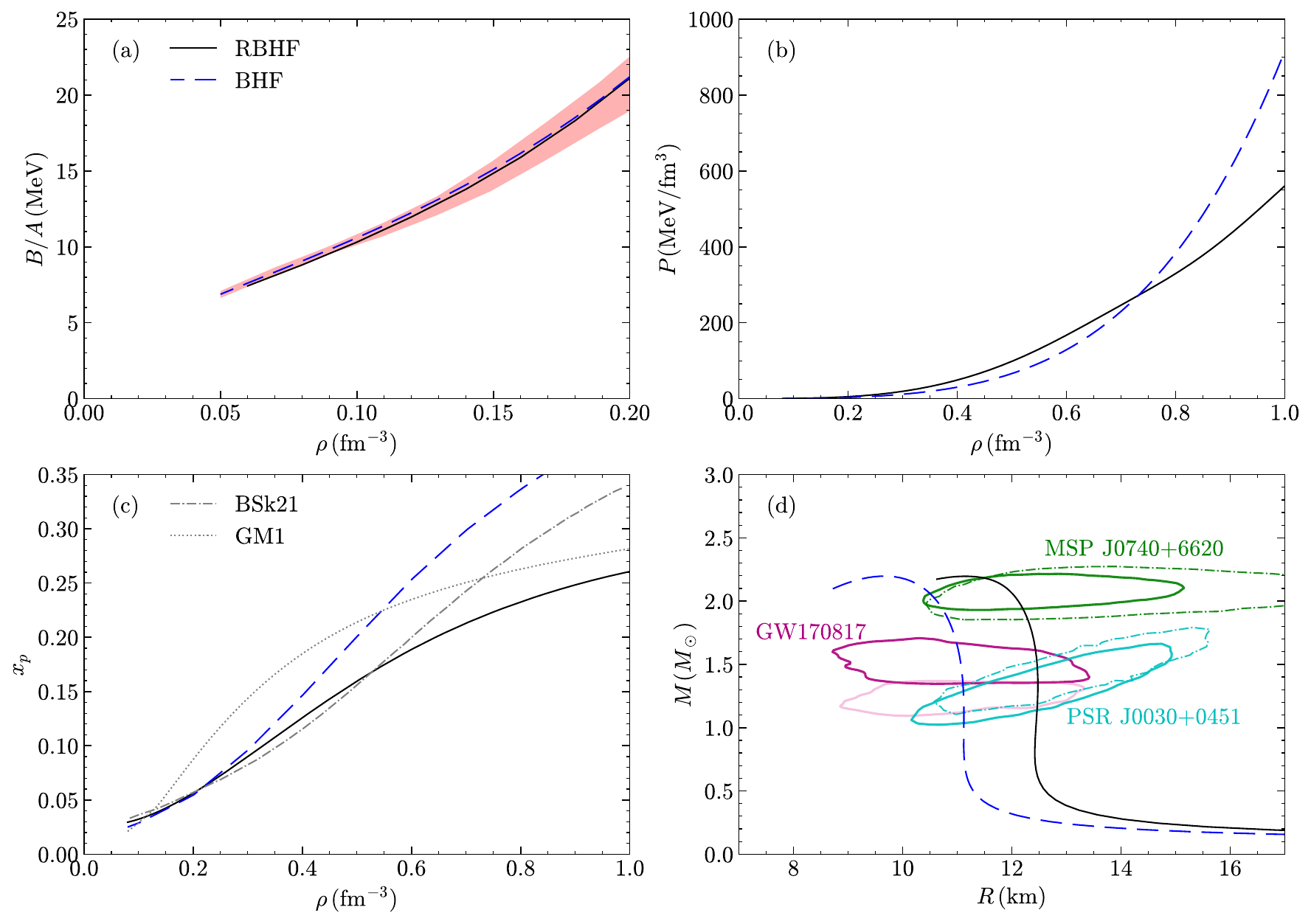}}
\caption{Energy per particle (panel a) of pure neutron matter are shown together with the EOS (panel b) and the proton fraction (panel c) of neutron star matter, as well as the mass-radius relation (panel d) for neutrons stars. 
The BHF results are from solving exactly the nuclear matter Bethe-Goldstone equation with the realistic Argonne $V_{18}$ including also a microscopic three-body force~\citep{2021PhRvC.103c4316S}.
The RBHF results are calculated with the latest high-precision, relativistic charge-dependent potentials, pvCD-Bonn A~\citep{2019ChPhC..43k4107W}.
In panel a, the shaded band represents the predictions with the chiral effective field theory reported in~\citet{2020PhRvL.125t2702D}, up to N3LO. In panel c, the proton fractions of GM1~\citep{1991PhRvL..67.2414G} and BSk21~\citep{2010PhRvC..82c5804G} are also included as representative examples considered in other studies~\citep{2013ApJ...764L..25H,2018MNRAS.475.5403A}.
The mass-radius constraints of the NICER mission for PSR J0030+0451~\citep{2019ApJ...887L..24M,2019ApJ...887L..21R} and MSP J0740+6620~\citep{2021arXiv210506980R,2021arXiv210506979M} are compared in panel d, along with the binary tidal deformability constraint from GW170817 by LIGO/Virgo~\citep{2017PhRvL.119p1101A,2018PhRvL.121p1101A}.
}
\label{f:EOS}
\end{figure*}

In Figure~\ref{f:EOS}, we present the binding energy of pure neutron matter, as well as the EOS and the proton fraction of neutron star matter for the BHF and RBHF models described above, together with the mass-radius relations of stable neutron stars from solving the TOV equations.
It is important to note here that the neutron star sequences such described, without any adjusted parameter, can, on the one hand, agree with the current constraints on the saturation properties of nuclear matter (as reported earlier in Table \ref{t:eos}),  as well as the results obtained for pure neutron matter within the chiral interactions at next-to-next-to-next-to-leading order (N3LO) level~\citep{2020PhRvL.125t2702D}, and on the other hand, meet the latest astrophysical mass and radius constraints from PSR J0030+0451~\citep{2019ApJ...887L..24M,2019ApJ...887L..21R} and MSP J0740+6620~\citep{2020NatAs...4...72C,2021ApJ...915L..12F,2021arXiv210506979M,2021arXiv210506980R}, as well as the binary tidal deformability constraint from GW170817~\citep{2017PhRvL.119p1101A,2018PhRvL.121p1101A}. 
Based on these EOSs and consistently determined stellar composition from the BHF and RBHF models, in the following we shall proceed to discuss the glitch observations of the Vela pulsar in the snowplow model of \citet{2011ApJ...743L..20P}. 

\begin{figure}
\centering
\includegraphics[scale=0.5]{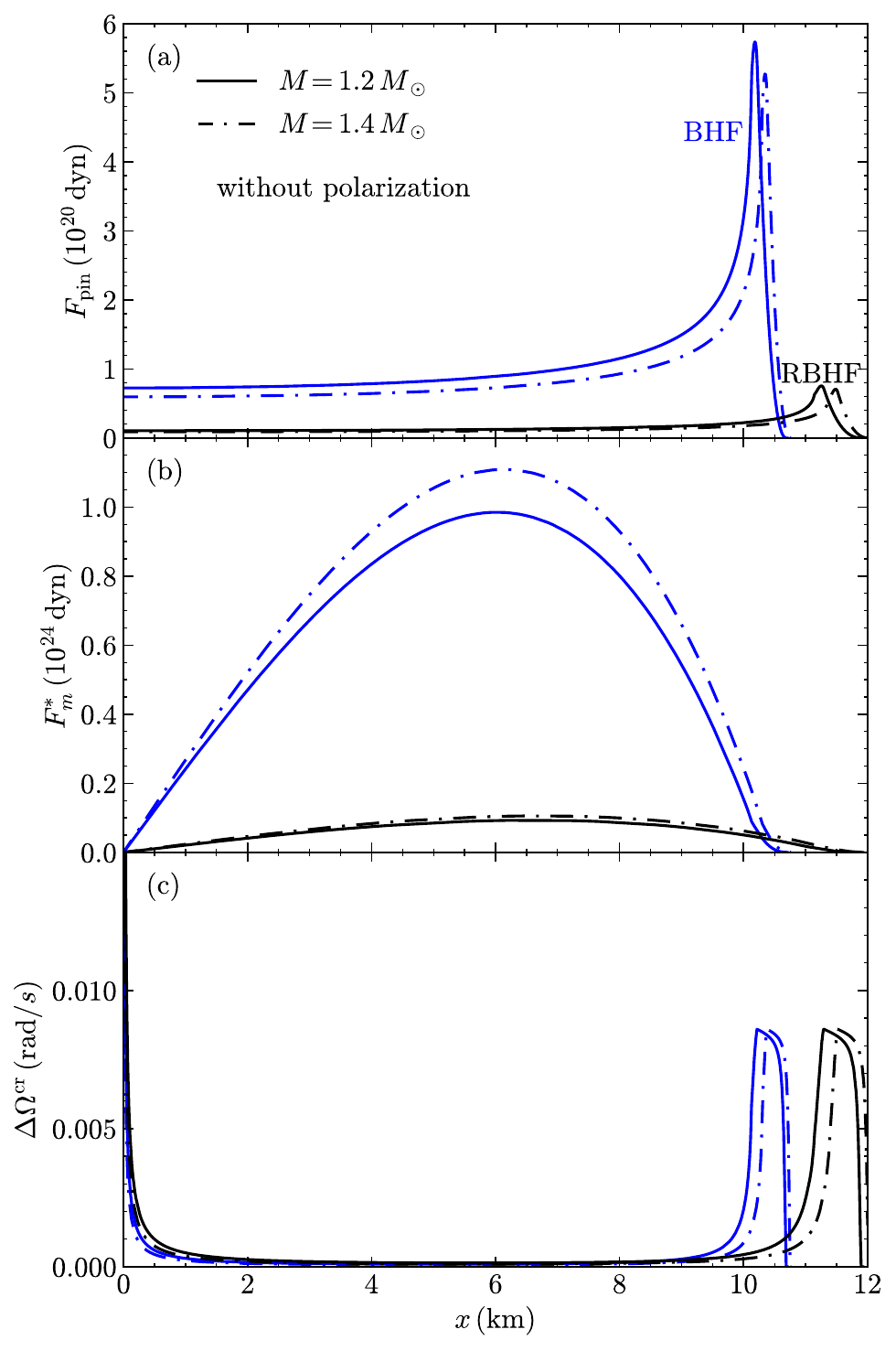}
\caption{Pinning force (panel a), Magnus force (panel b), and critical angular momentum lag (panel c) are shown as functions of the cylindrical radius (i.e., the distance of the vortex line to the rotational axis of the star), based on the EOS and composition of neutron stars determined from the BHF and RBHF models (see details in Section~\ref{sec:bhf}). 
The calculations are done without the including the polarization effect on the pairing gaps and for star mass of $M=1.2\Msun$ (solid lines) and $1.4\Msun$ (dash-dotted lines).
The peaks in the lower panel corresponds to $\Delta\Omega^{\rm cr}_{\rm
max}=\Delta\Omega^{\rm cr}(x_{\rm max})$, which are located in the star' inner crust, for example at density $\rho_{\rm max}=0.0544~\rm fm^{-3}~(\sim\negthickspace0.32\rho_0$) and $0.0543~\rm fm^{-3}~(\sim\negthickspace0.319\rho_0$) with the BHF EOS for $M=1.2\Msun$ and $1.4\Msun$, respectively.
}
\label{f:Fpin1}
\end{figure}

\begin{figure}
\centering
\includegraphics[scale=0.5]{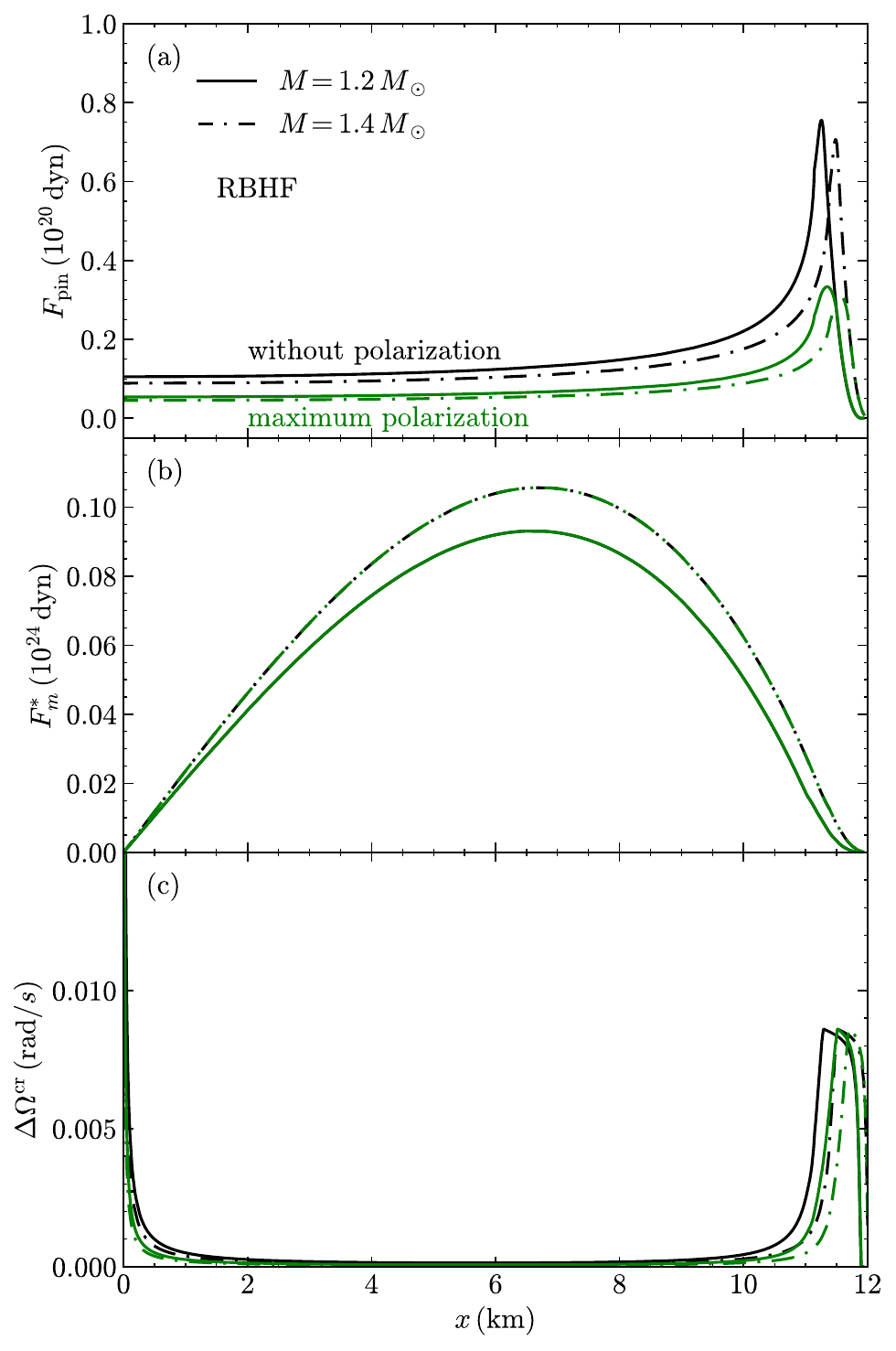}
\caption{Same with Figure~\ref{f:Fpin1} but for the results of the RBHF model with or without the including the polarization effect, shown in black and green curves, respectively.
}
\label{f:Fpin2}
\end{figure}

\section{Glitch size and short time-scale relaxation in the snowplow scenario}
\label{sec:step}

\subsection{Pinning force}

In the snowplow scenario~\citep{2011ApJ...743L..20P}, the
superfluid vorticity is assumed to accumulate in a thin sheet in
the inner crust and expected to be released simultaneously at some
cylindrical radius $x_{\rm max}$,
when the critical lag between the two
components reaches the maximum value $\Delta\Omega^{\rm cr}_{\rm
max}$.
The critical lag $\Delta\Omega^{\rm cr}(x)$ is
determined by the competition of the pinning force:
\begin{eqnarray}
F_{\rm pin}(x)=2\int_{0}^{l(x)/2}f_{\rm pin}[\rho(\sqrt{x^{2}+z^{2}})]dz
\end{eqnarray}
and the Magnus force:
\begin{eqnarray}
F_{\rm m}(x)=2\kappa x
\Delta\Omega(x)\int_{0}^{l(x)/2}\rho_{s}(\sqrt{x^{2}+z^{2}})dz ,
\end{eqnarray}
i.e., $\Delta\Omega^{\rm cr}(x)=F_{\rm pin}(x)/F_{\rm m}^{*}(x)$ with
$F_{\rm m}^{*}(x)=F_{\rm m}(x)/\Delta\Omega(x)$.
Here, the vortex line should be parallel to the
rotational axis of the star with a distance $x$.
$l(x)=2\sqrt{R_{\rm ic}^{2}-x^{2}}$ represents the length of the
vortex tube with the radius of inner crust $R_{\rm ic}$.
The constant $\kappa=\pi\hbar/m_{N}$ is the
quantum of circulation of a neutron fluid.
The function $\rho_{\rm s}(r)$ is the density of the superfluid component at radius
$r$.
We assume that the neutrons are superfluid throughout the star:
$\rho_{\rm s}=(1-x_{p})\rho$ with $x_{p}$ being the proton fraction calculated within BHF/RBHF (see Figure~\ref{f:EOS}).

To determine the critical lag, the pinning force per unit length $f_{\rm pin}(\rho)$, which acts on the vortex line stemming from its interaction with the lattice in the inner crust, should be evaluated from the pinning energy. Numerical simulations to calculate this quantity have been performed by considering the orientation of the bcc lattice~\citep{2012JPhCS.342a2004G}.
The results have shown that the order of magnitude of the maximum pinning force is approximately $10^{15}\rm~dyn~cm^{-1}$ and that the density where the pinning force reaches its maximum is in the inner crust density realm, i.e., from the neutron drip density to half of the nuclear saturation density $\rho_0$.
\citet{2012MNRAS.427.1089S} later simplified the function $f_{\rm pin}(\rho)$; i.e., the pinning force linearly increases up to the maximum magnitude and then linearly decreases. They also assumed that the pinning force would vanish at the boundary of the inner crust because the lattice exists only in the crust and that in the outer crust, there are no free neutrons to produce vortices.
The location of the maximum pining force crucially depends on the pairing gap. Yet, owing to the complexity of the nuclear medium polarization effects, referring to the self-energy (which affects the density of states for pairing) and the correction of the interaction vertex (which affects the pairing strength) \citep{1996PhLB.375,2003PhRvC.67}, the exact information of the pairing gap remains an open problem in nuclear physics.
However, different calculations indicate the polarization effects correspond to a reduction of the pairing gap calculated with a realistic interaction by a factor between two and three~\citep{2001lecture}. Therefore a reduction factor $\beta$, i.e., $\Delta=\Delta_{0}/\beta$, is introduced to account for the polarization effect in Ref.~\citep{2006PLB.640}. Where the gap $\Delta_{0}$ is calculated by using the free single particle energy with microscopic nucleon-nucleon potential (Argonne). The maximum polarization effect corresponds to $\beta=3$ with the maximal pairing gap of about $1$ MeV, while $\beta=1$ corresponds to nonpolarized effect with the maximal pairing gap of about $3$ MeV. This definition is adopted in~\citet{2012JPhCS.342a2004G,2012MNRAS.427.1089S} to evaluate the pining force. In the present calculation, we take the same $f_{\rm pin}(\rho)$ in two extreme configurations, i.e., the nonpolarized effect (indicated by $\beta=1$) and the maximum polarization effect (indicated by $\beta=3$),  as in ~\citet{2012MNRAS.427.1089S}. Also, the shape of $f_{\rm pin}(\rho)$ is adopted from~\citet{2012MNRAS.427.1089S}, whereas the magnitude of the pinning force in the both configuration is chosen to reproduce the max critical lag value $\Delta\Omega^{\rm
cr}_{\rm max}$ of the Vela-like pulsars.

Here, we follow \citet{2012MNRAS.427.1089S} to evaluate the critical lag.
In Figures~\ref{f:Fpin1} and~\ref{f:Fpin2}, we report the pinning force $F_{\rm pin}(x)$, the Magnus force $F^*_{\rm m}(x)$, and the critical lag $\Delta\Omega^{\rm cr}(x)$ by adopting the BHF/RBHF input in Section~\ref{sec:bhf}.
Figure~\ref{f:Fpin1} demonstrates the differences in the BHF and RBHF results in the case without polarization, while in Figure~\ref{f:Fpin2}, the results with or without the inclusion of polarization are compared in the RBHF case.
We see that there is no apparent difference in the shapes of the pinning force, the Magnus force, and the critical lag for different cases, except for their maximum values and the related locations.
The pinning force and the critical lag $\Delta\Omega^{\rm cr}(x)$ for $M=1.4\Msun$ are sharper than those for $M=1.2\Msun$ since a larger stellar mass leads to a more compressed stellar structure and a smaller $(x_{\rm max},R_{\rm ic})$ region.
Similarly, the softer BHF EOS results in a more compressed stellar structure and a smaller $(x_{\rm max},R_{\rm ic})$ region than the RBHF EOS.
When the polarization effect is considered, the density for the maximum pining force shifts to lower densities \citep{2012JPhCS.342a2004G,2012MNRAS.427.1089S}. Accordingly, the region $(x_{\rm max},R_{\rm ic})$ is compressed.
Since the angular momentum reservoir is assumed to accumulate in the thin sheet from $x_{\rm max}$ to $R_{\rm ic}$ in the snowplow model, a smaller $(x_{\rm max},R_{\rm ic})$ region should correspond to less available transferred angular momentum, as will be expressed in Equation (\ref{eq23}) in Section~\ref{sec:stepB}.
In addition, because the Magnus force $F^*_{\rm m}(x)$ is free of polarization, $F^*_{\rm m}(x)$ in the cases with or without polarization is identical.

\subsection{Frequency jump and fast-decaying component of the glitch}
\label{sec:stepB}

Immediately before a glitch, a lag of $\Delta\Omega^{\rm cr}_{\rm
max}$ will have been created at $x=x_{\rm max}$. For an axial symmetry
system, the angular velocity $\Omega_{\rm n}(x)$ of the superfluid
component of the star is in fact proportional to the number $N(x)$
of vortices enclosed in a cylindrical region with radius $x$, and
can be expressed as follows:
\begin{eqnarray}
\Omega_{\rm n}(x)=\frac{\kappa N(x)}{2\pi x^{2}}\ .
\end{eqnarray}
One can use this expression to evaluate the number of vortices stored at the peak just before a glitch, i.e.:
\begin{eqnarray}
N_{\upsilon}=\frac{2\pi}{\kappa}x_{\rm max}^{2}\Delta\Omega^{\rm
cr}_{\rm max}\ .
\label{eq21}
\end{eqnarray}
Moreover, this max critical lag value $\Delta\Omega^{\rm cr}_{\rm
max}$ can also be evaluated by the time interval between two
glitches $\Delta t_{\rm g}$ and the spin-down rate:
\begin{eqnarray}
\Delta\Omega^{\rm cr}_{\rm max}=\Delta t_{\rm g}|\dot{\Omega_{\rm p}}| \ .
\end{eqnarray}

Due to the particular shape of $\Delta\Omega^{\rm cr}(x)$ in
Figures~\ref{f:Fpin1} and~\ref{f:Fpin2}, one can assume that just before a glitch, the excess
vortices in the region $x>x_{\rm max}$ have been entirely removed by
the Magnus force, and consequently only the $N_{\upsilon}$
vortices respond to the angular momentum transfer.
Following the procedure in the ``snowplow'' scenario~\citep{2012MNRAS.427.1089S}, we employ $dL_{\rm g}=\Omega_{\rm n}(x)dI_{\rm n}$ to evaluate
this angular momentum transfer. One should note that only the $N_{\upsilon}$ vortices at the peak of the pinning potential ought to be considered in the calculation. After performing the integration on the
cylindrical region $x_{\rm max}<x<R_{\rm ic}$, the requested angular momentum transfer can be obtained as follows:
\begin{eqnarray}
\Delta L_{\rm g}=2\kappa
N_{\upsilon}\int_{x_{\rm max}}^{R_{\rm ic}}xdx\int_{0}^{l(x)/2}\rho_{\rm s}(\sqrt{x^{2}+z^{2}})dz\ .
\label{eq23}
\end{eqnarray}

As mentioned above, during the glitch, only a fraction of the core neutrons might be coupled
to the charged component.
The parameter $Y_{\rm gl}$ is introduced to describe the fraction that contributes to the charged component of the star at the time of glitch~\citep{2011ApJ...743L..20P, 2012MNRAS.427.1089S,2013ApJ...764L..25H,2015MNRAS.449.3559H}.
$Y_{\rm gl}$ is related to the trigger mechanism of glitches and cannot be determined from the current simple ``bulk'' model.
Presently, we do not have a good estimation from the hydrodynamic simulation, and we treat it as a free parameter to be determined by reproducing the observed glitch size.
$Y_{\rm gl}=1$ means that the core superfluidity is fully coupled to the charged component during a glitch.

\begin{figure}
\centering
\resizebox*{0.45\textwidth}{0.35\textheight}
{\includegraphics{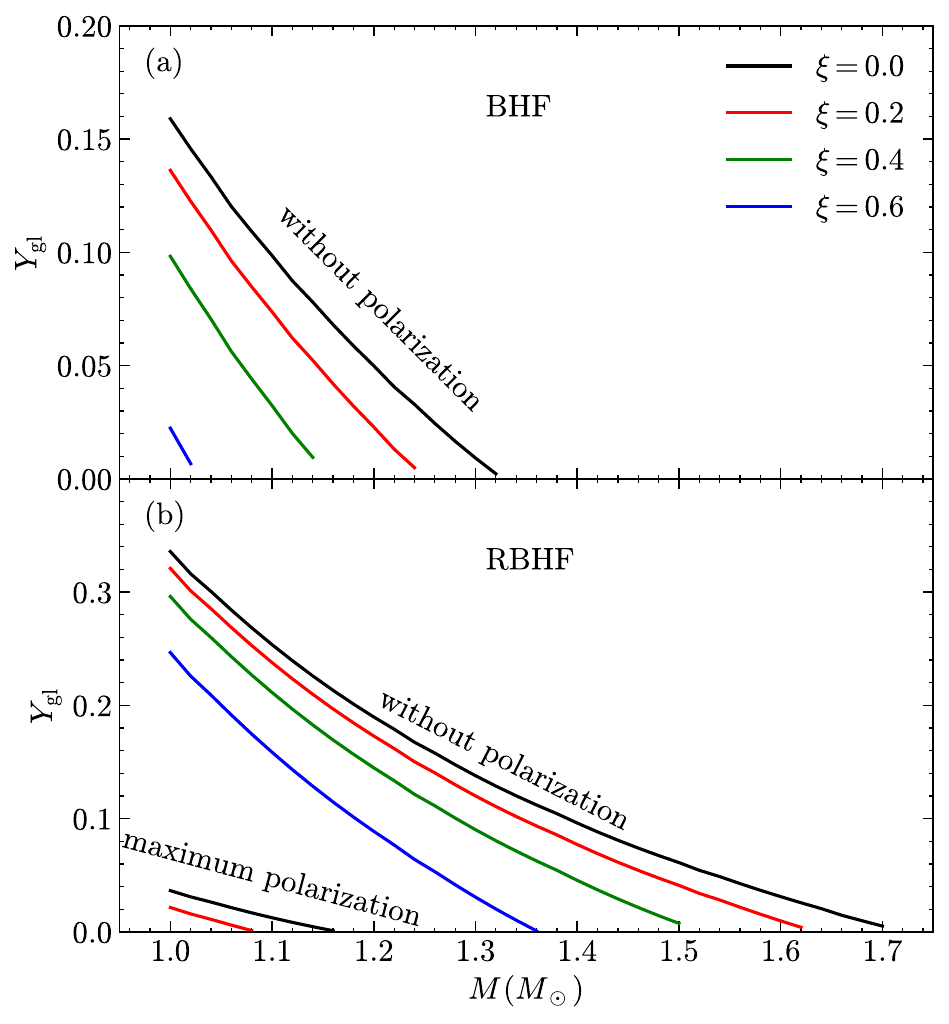}
}
\caption{Fraction of the coupled vorticity at the glitch $Y_{\rm gl}$ as function of the stellar mass for varying values of the fraction of pinned vorticity in the core ($\xi$). 
The calculations are done for both the BHF (panel a) and RBHF (panel b) model. 
And in panel (b) of the RBHF EOS, the results including the maximum polarization are shown together with those without polarization, while unphysical (negative) values for $Y_{\rm gl}$ are not reported in panel (a) of the BHF EOS case.
In fact, the result that there is no positive $Y_{\rm gl}$ for $M> \Msun$ in such a case suggests that the stored angular momentum of the vortices cannot support the magnitude of the 2000 Vela glitch in such configuration.}
\label{f:ygl}
\end{figure}

\begin{figure*}
\centering
\resizebox*{0.98\textwidth}{0.5\textheight}
{\includegraphics{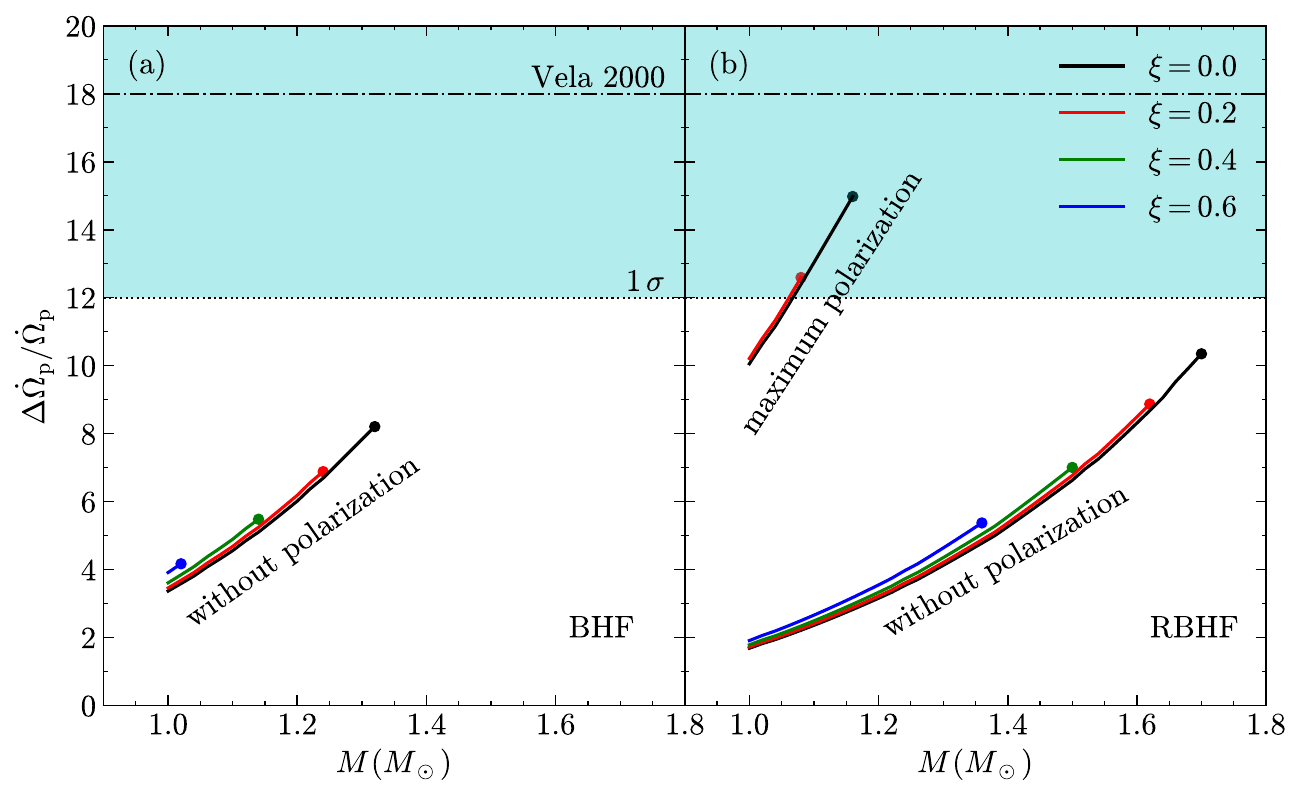}
}
\caption{Observed step in spin-down rate on short time-scales, confronted with the theoretical results from the microscopic neutron star EOSs from the BHF model (panel a) and the RBHF model (panel b).
The horizontal line represents the measured value for the Vela 2000 glitch, together with the $1\sigma$ deviation.
Note that the calculations are done without including the non-polarized effect on the pairing gap ($\beta=1$) in the BHF case, due to that no physical paramter can be found in the polarized-gap case, as explained along with Figure~\ref{f:ygl}. 
}
\label{fig:dww}
\end{figure*}

Defining $Q$ as the superfluid fraction of the total moment of inertia:
\begin{eqnarray}
Q=\frac{I_{\rm n}}{I_{\rm tot}}=\frac{\int_{0}^{R}r^{4}(1-x_{p}(\rho))\rho(r)dr}{\int_{0}^{R}r^{4}\rho(r)dr}\ ,
\end{eqnarray}
the observed jump in the angular velocity is thus:
\begin{eqnarray}
\Delta\Omega_{\rm p}=\frac{\Delta L_{\rm g}}{I_{\rm tot}[1-Q(1-Y_{\rm gl})]}\ ,
\end{eqnarray}
with the total moment of inertia $I_{\rm tot}=I_{\rm n}+I_{\rm p}$.
If the recoupling of the rest of the core superfluid is responsible for the relative angular deceleration of the charged
component immediately after the glitch, from angular
momentum conservation $\Delta L_{\rm g}=\Delta(I_{\rm p}\Omega_{\rm p})=0$,
the relative deceleration of the crust immediately after the
glitch follows:
\begin{eqnarray}
\frac{\Delta
\dot{\Omega_{\rm p}}}{\dot{\Omega_{\rm p}}}=\frac{Q(1-Y_{\rm gl})}{1-Q(1-Y_{\rm gl})}\ ,
\end{eqnarray}
where $\dot{\Omega_{\rm p}}$ is the steady state pre-glitch spin-down rate.

In addition to a partially coupled core, we also consider that the protons in neutron star cores possibly form a type II superconductor in which the magnetic field is carried by flux tubes; thus, the strong interaction between the flux tubes and the neutron rotational vortices could impede the vortex motion and effectively pin some of the neutron vorticity in the core~\citep{1998ApJ...492..267R,2009MNRAS.400.1859S}.
\citet{2013ApJ...764L..25H} proposed a parameter $\xi$, which is defined as the fraction of pinned vorticity in the core by the flux tubes, to account for this effect.
Namely, this $\xi$ vorticity does not contribute to the angular momentum transfer, i.e., the total number of vortices stored at $x_{\rm max}$ just before a glitch is instead:
\begin{eqnarray}
N_{\upsilon}=(1-\xi)\frac{2\pi}{\kappa}x_{\rm max}^{2}\Delta\Omega^{\rm
cr}_{\rm max}\ .
\end{eqnarray}
One should note that the effects of the flux tube are not included completely, especially, the effect of the flux tube on the pinning force, which can influence all stages of the glitch dynamics~\citep{1998ApJ...492..267R,2014ApJ..788L..11G,2020MNRAS.496.2506G, 2018APJ..865..23, 2020MNRAS.493.L98}, is omitted.
Substituting the above expression of $N_{\upsilon}$ for Equation (\ref{eq21}), one can obtain the angular momentum exchanged during the glitch. Accordingly, the coupled fraction of superfluid $Y_{\rm gl}$ can be derived by fitting the size of a glitch, $\Delta\Omega_{\rm p}$:
\begin{eqnarray}
Y_{\rm gl}=\frac{1}{Q(1-\xi)}\Bigg[\frac{\Delta L_{\rm g}}{\Delta\Omega_{\rm p}I_{\rm tot}}+Q-1\Bigg].
\end{eqnarray}
The instantaneous step in the frequency derivative then follows:
\begin{eqnarray}
\frac{\Delta
\dot{\Omega_{\rm p}}}{\dot{\Omega_{\rm p}}}=\frac{Q(1-\xi)(1-Y_{\rm gl})}{1-Q[1-Y_{\rm gl}(1-\xi)]}\ .
\end{eqnarray}
For the 2000 January glitch from the Vela pulsar, $\Delta\Omega_{\rm p}=2.2\times10^{-4}~\rm rad~s^{-1}$.
The 2000 Vela glitch also yielded the first measurement of the fractional change in the spin-down rate on a short time scale of approximately one minute, ${\Delta\dot{\Omega}_{\rm p}}/{\dot{\Omega}_{\rm p}}=18\pm6$ (1$\sigma$ error)~\citep{2002ApJ...564L..85D}.

In Figure~\ref{f:ygl}, the obtained $Y_{\rm gl}$ in the present study is shown as a function of the stellar mass under various choices of $\xi$ for the BHF and RBHF models.
It is seen that $Y_{\rm gl} \lesssim 15\%$ for BHF and $Y_{\rm gl}\lesssim 30\%$ for RBHF in the case of an unpolarized gap.
A maximally polarized gap would result in a much loosely coupled core, with $Y_{\rm gl}\lesssim 4\%$ for RBHF.

In Figure~\ref{fig:dww}, we confront a consistent and microscopic neutron star description with the observed short time scale relaxation immediately after the 2000 Vela glitch.
We see that consistent with previous works~\citep{2011ApJ...743L..20P,2012MNRAS.427.1089S,2013ApJ...764L..25H}, the pinned core vertices are disfavoured by the observed ${\Delta\dot{\Omega}_{\rm p}}/{\dot{\Omega}_{\rm p}}$; hence, most of the vorticity in the core should be free.
One main point here is that when the pairing gap is considered to be not polarized, one cannot fit the step in the frequency derivative at the $1\sigma$ level even in the wholly unpinned case $\xi=0$.
This is the case for both BHF and RBHF models.
It is only barely achievable at the $1\sigma$ level if less than $\sim\negthickspace30\%$ of the vorticity in the core is pinned with for a maximally polarized gap in the stiffer RBHF case because the RBHF EOS allows a large stellar radius ($\sim\negthickspace12.5\km$ compared to $\sim\negthickspace11.1\km$ with BHF; see Figure~\ref{f:EOS}) and the correspondingly thicker accumulating sheet.
One should realize, however, that the above results are obtained without considering the entrainment effect previously mentioned.
The inclusion of the entrainment effect would further worsen the situation. For example, reducing the available angular momentum by a factor of $\langle m_n^* \rangle/m_n\sim\negthickspace1.35$ would make it impossible to reconcile with the observations even for RBHF, especially that the actual neutron superfluid should be more strongly entrained by the crust, with $\langle m_n^* \rangle/m_n$ greater than $\sim\negthickspace1.5$~\citep{2005NuPhA.747..109C,2012PhRvC..85c5801C,2017PhRvL.119f2701W,2017JLTP..189..328C}.
The current framework also provides stringent constraints on the mass of the Vela pulsar, i.e., less than $1.16\Msun$, as seen in Figure~\ref{fig:dww} (b), if both the observed glitch size and the post-glitch jump are to be reproduced.
The inferred mass of the Vela pulsar is not only much smaller than previous values, either from the
largest observed glitch in a pulsar~\citep{2017NatAs...1E.134P} or from combining X-ray with radio data~\citep{2015SciA....1E0578H},
but also approaches that of the possible lightest measured neutron star, the Pulsar J0453+1559 companion of mass $1.174\pm0.004\Msun$~\citep{2015ApJ...812..143M}.
How such light neutron stars are formed and their implications on progenitors and binary stellar evolution are presently open problems.
Finally, our calculations on ${\Delta\dot{\Omega}_{\rm p}}/{\dot{\Omega}_{\rm p}}$ are based on a strong assumption that the post-glitch deceleration is caused entirely by the increase in the effective moment of inertia of the star; however, the crust-core mutual friction is suggested to play an important role~\citep{2015MNRAS.454.4400N}. Future studies are planned by incorporating both effects to reproduce the glitch magnitudes and the post-glitch rotational evolution.

\section{Summary and discussions}\label{sec:result}

The glitch observation of Vela-like pulsars (especially the details on short-term morphology) may provide a test of the standard superfluid glitch theory.
To do so, it is crucial to make use of consistent and microscopic descriptions of the matter state and the structure of neutron stars.
In the present study, our input star properties are derived from the microscopic BHF/RBHF theory, including a consistent determination of the star EOS and composition.
The two representative EOSs (without any adjusted parameter) have been tested from laboratory nuclear experiments and the latest astrophysical observations and thus are regarded as reliable at $\sim\negthickspace 0.5\negthickspace-\negthickspace3\,\rho_0$ times the nuclear saturation density, which is crucial to determine the structure of the crust and therefore the pinning vortices.

Within the particular EOS models we use, we found that, although the snowplow scenario of the simple two-component model has no problem matching the amplitudes of the glitches from the Vela pulsar, it could hardly explain the one available observation of short-term relaxation from the January 2000 glitch.
Additionally, the Vela mass is unexpectedly small.
These results thus challenge to some extent the present understanding of the simple two-component framework of superfluid glitch theory, and strongly support the conclusions that the mutual friction may closely relate to post-glitch recoveries.
Extended studies on the post-glitch rotational evolution with a large sample of EOSs will be reported in a separate work. Further efforts should also be made to construct a unified theory able to describe on a microscopic level the complete structure of neutron stars from the outer crust to the core.
Upcoming higher time resolution measurements with post-glitch relaxation should provide more insight into these issues, especially because some real-time automated glitch detection pipelines have been presented~\citep{2021MNRAS.505.5488S}, which can measure the postglitch recovery phase at a high cadence.

On the other hand, there is increasing evidence of radiative changes accompanying glitch events in magnetars~\citep{2008ApJ...673.1044D,2014ApJ...784...37D,2017ARA&A..55..261K} and radio pulsars~\citep{2011MNRAS.411.1917W,2013MNRAS.432.3080K,2018MNRAS.478L..24K,2018Natur.556..219P}.
For example, there are changes in the spin-down state, emission mode changes and narrowing of the pulse associated with glitch activity in PSR B2035+36~\citep{2018MNRAS.478L..24K}.
\citet{2018Natur.556..219P} detected sudden changes in the pulse shape coincident with the glitch event of the Vela pulsar.
In anomalous X-ray pulsar 1E 2259+586, a glitch was observed preceding an X-ray burst~\citep{2004ApJ...605..378W}.
One of the brightest radioquiet gamma-ray pulsars, PSR J2021+4026, showed a sudden decrease in gamma-ray emission at the glitch~\citep{2017ApJ...842...53Z}.
Recently, \citet{2020NatAs...4..511F} reported an association between the glitch and polarization change of the Crab pulsar. They proposed that the sudden spin-up of the crust may cause a change in the configuration of magnetic fields threaded in it and consequently in the co-rotating magnetosphere.

Therefore, although no variation could be seen either before or after the 2000 Vela glitch in profile shape, magnitude, or polarization in the radio observations on any timescale,
for future studies aiming for a thorough understanding of various observed glitch behaviours,
a coherent picture involving stellar interiors and their magnetosphere may be necessary.
Further information will soon be obtained from high-accuracy facilities, e.g., FAST or SKA, with the help of multiband collaborative observations.
We expect this information to give additional constraints to the viability of the superfluid model or to identify the glitch (together with preceding slowdown) that arises by other means, such as magnetospheric activities, hydrodynamic instabilities and turbulence, or starquakes.

\section*{Acknowledgements}
We are thankful to an anonymous referee for valuable comments and suggestions. 
We would also like to thank J.-M. Dong, M.-Y. Ge, P. Wang, J.-P. Yuan, X. Zhou, B. Haskell for helpful discussions and J. N. Hu, C. C Wang for providing us the RBHF EOS.
The work is supported by National SKA Program of China (No.~2020SKA0120300), the Strategic Priority Research Program of Chinese Academy of Sciences (No. XDB34000000), the National Natural Science Foundation of China (Grant No.~11873040), the science research grants from the China Manned Space Project (No. CMS-CSST-2021-B11), the Youth Innovation Promotion Association of Chinese Academy of Sciences (No. Y2021414), and the Youth Innovation Fund of Xiamen (No. 3502Z20206061).

\end{document}